\documentclass[aps,pra,twocolumn]{revtex4}  
\usepackage{graphicx}  
\usepackage{float}
\usepackage{subcaption}
\usepackage{dcolumn}   
\usepackage{bm}        
\usepackage{amssymb}   
\usepackage{float}
\usepackage{enumerate} 
\usepackage{amsmath}
\usepackage{xcolor}
\usepackage{bm}
\usepackage[utf8]{inputenc}
\usepackage[T1]{fontenc}
\usepackage[spanish]{babel}
\usepackage{lipsum}   
\usepackage{esvect}

\usepackage{hyperref}
\hypersetup{
	colorlinks = false,
	linkbordercolor = {green}
}

\begin{document}



\title{Quantum Walk and Quantum Rings. Effects of a Magnetic Field applied and construction of Quantum Walk by a Moire pattern }

\author{C\'esar \ Alonso $-$ Lobo \ \  ,  \ \ Manuel \ Martínez $-$ Quesada *}


\begin{abstract}
	
\begin{center}	
	\textit{* Departamento de \'Optica , Optometr\'ia y Ciencias de la Visi\'on, Universidad de Valencia , Dr. Moliner 50 , 46100-Burjassot}
\end{center}

Quantum Rings have been simulated so far in many ways, but in this work a new aproximation is deemed. We use particles without angular momentum and several spectra, for different geometric settings, are gotten. These spectra depends on K, not on L. The application of a magnetic field is also analysed, and an Aharonov-Bohm kind of effect is observed. Some new math technique is introduced as well. We think that with this new approach the spectrum matrix can be treated in a distinct way, and in such a manner is how we get to handle the Moire pattern spectrum. Furthermore, we show that two electrons quantum walking by some kind of double concentric Quantum Rings and those very same electrons quantum walking by a Moire circunference could indeed have, if not for some parameters, the same spectrum. So they might somehow display similar behaviour and similar properties
\end{abstract}

\maketitle

 

\section{Introduction}

This paper is based on the Quantum Walk \cite{QW Book} \cite{Navarrete} (\textbf{QW} hereafter) which is a mechanism of quantum propagation over a grid. In such a framework at first you toss a coin, and depending on the result you get you move towards the right or to the left. 
With this very simple mechanism not only the movement of a particle over a line has been described, \cite{QW}, but also on a honeycomb grid \cite{HoneyComb} ( to model an electron moving by a graphene-like structure ), or its walk over a 2-D lattice \cite{QW 2D} that can be  understood as the propagation of two particles as well, and  those particles can be made to interact via a Coulomb potential for instance, and hence a molecule appears \cite{QW Coulomb}.
Besides, several simulations have been studied related with the \textbf{QW} of a non free particle, but with a particle under the action of some external electric field applied \cite{QW E} or under the influence of an external magnetic field \cite{QW B} \cite{QW B Fractal}. How you can introduce any magnetic field is excellently well explained in \cite{Feynman B} and in the section 2 of this paper as well. Furthermore several papers have been published concerning the decoherence that affects the quantum nature of the particle that propagates, and how to avoid such an undesirable effect \cite {QW Decoherence} because if decoherence happens you lose the quantum speed up of the process. And of course not only theoretical works have been done, a lot of papers focused on the experimental applications of this idea have been realised \cite{QW experimental} and \cite{QW exp 2}
\\

With this background we have managed to define an operator that according to the aforementioned theoretical framework describes the propagation of two 1/2 spin particles over two concentric rings (the extrapolation to higher spin particles is straighforward once you have learnt what we do here). Such an arrangement is called Quantum Rings in the bibliography (\textbf{QR} hereafter) \cite{QR Book}, \cite{QR Review}. In our study these two rings connect from time to time between them and permit the transit of the particle from one of them to the another one. This transit pretends to mimic the quantum tunnel effect that is present in the bibliography related to the \textbf{QR}. These other papers \cite{QR 1}, \cite{QR 2}, \cite{QR 3} model this propagation with the aid of the Schr\"{o}dinguer equation, although they use the angular momentum of the particle as a degree of freedom \cite{Angular Momentum} and we don't even consider that property of the particles we work with. That's why we don't find an accumulation of the higher momentum particles in the outer ring. Our approach is quite different in this way
\\

We also consider the introduction of some magnetic field (\textbf{B} hereafter) , and it leads to a displacement of the spectrum. All the frequencies are shifted towards the right by the same amount, and this amount has to do with the \textbf{B} applied. This is an  \textbf{Aharonov-Bohm} \cite{QW A-B} kind of effect appearing in our model because the \textbf{B} itself does not act on the particles. Indeed, it is well known that not only the magnetic fields affects the phase of the wave function of the moving particle, but also the potential vector A itself introduces some changes. So the variations introduced by the external \textbf{B} applied shift the phase of the amplitudes that a brief time after interfere, and therefore change the interference pattern. The effects of a \textbf{B} applied directly on a particle that moves on a 2-D lattice have been already studied in \cite{QW B Fractal}. But the way how the B applied acts on the particles changes in our model. And this fact leads us to a different effect
\\

Finally we have built an operator that models the \textbf{QW} of two particles by an overlapping of rings. We are now facing a ``Moire ring''. We try to get an understanding of what happens in such a propagation because it is one of the most naive models related to the superconductivity found experimentally a few years ago at M.I.T. \cite{MIT}. In the aforementioned experiment two lattices were overlapped in such a way that sometimes a propagation without almost any resistence was achieved.
\\

This paper is organised as follows. In \textit{Section 2} we introduce the operator that shapes the propagation of two quantum particles over two concentric rings. We also study in this section the effect of the insertion of a magnetic field B, and we explain with a lot of detail how the phase changes when you introduce such a field. And finally in \textit{Section 3} a kind of mathematical curiosity is introduced that enables us to attack the Moire pattern problem, which is more deeply developed in \textit{Section 4}

\section{QR operator via QW}

In the 1-D \textbf{QW} of an electron the evolution is as follows: First, on the wave function acts a coin operator $\hat{C}$ represented as an SU(2) matrix that mixes the two amplitudes of probability, and secondly acts the displacement operator defined as $\hat{W} = |x+1><x| \otimes |U><U| + |x-1><x| \otimes |D><D|$. Here, in the double \textbf{QR} system, the evolution is analogous. But now there are two 1/2 spin particles that propagates around two concentric rings in such a way that they only interfere in some places. The particle A ,so to speak, moves around the ring A, and the particle B does so around the ring B. This scheme may be better understood with the picture shown in figure \textbf{1}. In this drawing two concentric five nodes rings are displayed with one point where they connect to each other. These connections symbolize leaps from one ring to another, in a similar way to how some tunnel effect might occur. The number of leaps depend on the least common multiple (lcm) of the parameters \textit{a} and \textit{b} in such a way that, changing the values of  \textit{a} and \textit{b} the tunnel effect can be modeled to happen with more or less frequency.  The reason why we assume this definition will be made clear at the end of the paper
\\

\begin{figure}
	\includegraphics[width=8cm]{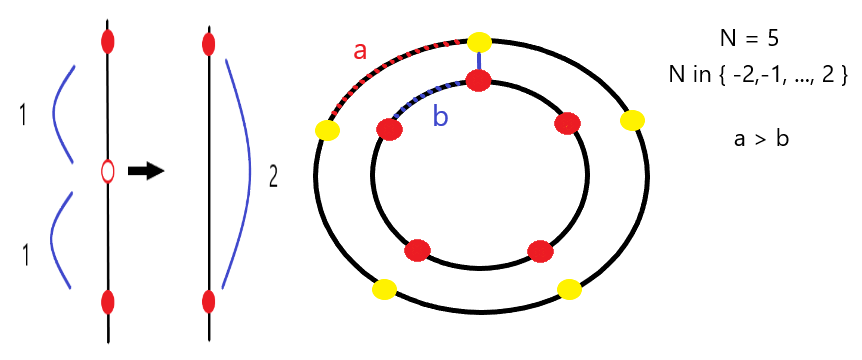}
	\caption{One possible example of double concentric quantum rings. Both rings have the same number of sites. The parameters a \& b take their values in function of 1 = elementary leap in some reference lattice. So a \& b $\in \Re$ in general }
	\medskip
	\includegraphics[width=8cm]{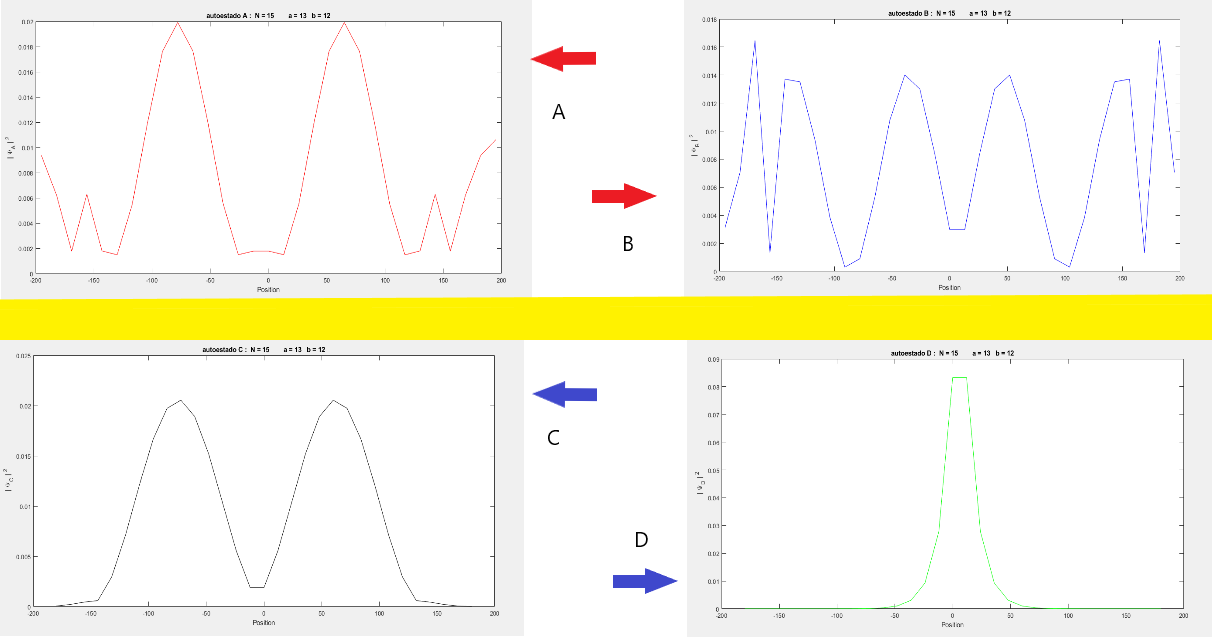}
	\caption{In this figure four instances of distribution of probability for different eigenstates are presented . $L_{1} = (2 N +1) \times a$ for eigenstates A and B, that propagates by the outer ring, the one that has a bigger radius. And, otherwise $L_{2} = (2 N +1) \times b $ for eigenstates C and D that goes by the inner one, with a smaller radius. See figure \textbf{6.1} in the left for further details }
\end{figure}

%
%
%

\vspace{2mm}

To achieve this movement by the two concentric rings the next coin operator is defined. It represents particle 1 with amplitudes A and B that goes clockwise and counter-clockwise around the outer ring and particle 2 with amplitudes C and D that moves by the inner ring. Also in some cases they interfere and hence you need a SU(4) coin operator instead, you need an operator that mixes the four probability amplitudes in the junction points. So, in conclusion, the operator $\hat{C}$ sometimes is a direct sum of two independent SU(2) coins $\hat{C_{1}}$ and $\hat{C_{2}}$ each of them acting on a different particle driven around one and only one of the two rings, and in case of connection between the two rings $\hat{C}$ is a SU(4) matrix that acts on both particles and mixes up the four amplitudes.
\\

\begin{widetext}
	
\begin{equation*}
 \mathbf{\hat{C}} =
\begin{pmatrix}  
	& \textcolor{blue}{cos(\Theta_{n})} * cos(\Theta_{1}) & \textcolor{blue}{cos(\Theta_{n})} *sin(\Theta_{1}) & \textcolor{blue}{sin(\Theta_{n})} * cos(\Theta_{1})  & \textcolor{blue}{sin(\Theta_{n})} * sin(\Theta_{1}) \\
	& \textcolor{blue}{cos(\Theta_{n})} * -sin(\Theta_{1})  & \textcolor{blue}{cos(\Theta_{n})} * cos(\Theta_{1}) & \textcolor{blue}{sin(\Theta_{n})} * - sin(\Theta_{1})  &   \textcolor{blue}{sin(\Theta_{n})} * cos(\Theta_{1}) \\
	& \textcolor{blue}{- sin(\Theta_{n})} * cos(\Theta_{1})  & \textcolor{blue}{- sin(\Theta_{n})} * sin(\Theta_{1})    & \textcolor{blue}{cos(\Theta_{n})} * cos(\Theta_{1})   & \textcolor{blue}{cos(\Theta_{n})} * sin(\Theta_{1})   \\
	&\textcolor{blue}{- sin(\Theta_{n})} * - sin(\Theta_{1})  & \textcolor{blue}{- sin(\Theta_{n})} * cos(\Theta_{1})   & \textcolor{blue}{cos(\Theta_{n})} * - sin(\Theta_{1})   &\textcolor{blue}{cos(\Theta_{n})} * cos(\Theta_{1})    \\
  \end{pmatrix}
\end{equation*}

\end{widetext}

In the previous operator $\Theta_{n}$ represents this angle :
$\Theta_{n} =$ $\Theta_{2} \ * \ ( \delta_{mod(n*a , lcm)}^{0} \ + \ \delta_{mod(n*b , lcm)}^{0} \ - \ \delta_{mod(n*a , lcm)}^{0} * \delta_{mod(n*b , lcm)}^{0} )$ where the delta function is a Cronecker delta. In this manner $\Theta_{n} = \Theta_{2} $ in the junction points, while $ \Theta_{n} = 0$ otherwise. For example if the values a = 9 , b = 7 and N = 5 are chosen, since the rings are placed as shown in figure \textbf{1}, the intersection between the two rings happens only in the position $0$ of both rings. But if the values a = 6 , b = 2 and N = 7 are fixed, both rings connect each other at any point and, in this case, an usual \textbf{QW} with a SU(4) coin is expected to be observed
\\

On the other hand  the walk operator is defined as follows.

$\hat{W} = | na +a, A> <na ,A| \ + \ | na - a ,B> <na,B| \ + \ |nb+ b, C> <nb,C| \ + \ |nb - b, D > <nb,D| $

Now the spectrum of the operator $\hat{U} = \hat{W} \otimes \hat{C}$ can be obtained in the following way:

\[
\begin{bmatrix} A_{n a}(t+1) \\ B_{n a}(t+1) \\ C_{n b}(t+1) \\ D_{n b}(t+1)  \end{bmatrix}
=\hat{W}
\begin{bmatrix}
c_{n} c_{1} &  c_{n} s_{1}& s_{n} c_{1}& s_{n} s_{1}\\
- c_{n} s_{1} & c_{n} c_{1} & - s_{n} s_{1} & s_{n} c_{1} \\
- s_{n} c_{1} & - s_{n} s_{1} & c_{n} c_{1} &  c_{n} s_{1}\\
s_{n} s_{1} & - s_{n} c_{1} & -  c_{n} s_{1} & c_{n} c_{1} 
\end{bmatrix}
\begin{bmatrix} A_{n a}(t) \\ B_{n a}(t) \\ C_{n b}(t) \\ D_{n b}(t)  \end{bmatrix}
\]

\vspace{2mm}

From this matrix equation is obvious to get the next set of equations that links the probability amplitudes in \textbf{t} with those amplitudes in the time \textbf{t+1}. We proceed as has already been done before in \cite{QW Coulomb}

\begin{itemize}
	
	\item	$\mathbf{1}$
	
	\vspace{2mm}
	
	$A_{n a}(t+1) \ \mathbf{|A, na >} = \ \mathbf{\hat{W}} \ (A_{na}(t) \ c_{n} c_{1} \ + \ B_{na}(t) \ c_{n} s_{1} + C_{n b}(t) \ s_{n} c_{1} + D_{n b}(t) \ s_{n} s_{1}) \ \mathbf{|A, na >} $
	
	\vspace{2mm}
	
	$B_{n a}(t+1) \ \mathbf{|B, na >} = \ \mathbf{\hat{W}} \ (- A_{na}(t) c_{n} s_{1} \ + \ B_{na}(t) \ c_{n} c_{1} - C_{n b}(t)  s_{n} s_{1} + D_{n b}(t) \ s_{n} c_{1}) \ \mathbf{|B, na >} $
	
	\vspace{2mm}
	
	$C_{n b}(t+1) \ \mathbf{|C, nb >} = \ \mathbf{\hat{W}} \ (- A_{na}(t) s_{n} c_{1} \ - \ B_{na}(t) \ s_{n} s_{1} + C_{n b}(t)  c_{n} c_{1} + D_{n b}(t) \ c_{n} s_{1}) \ \mathbf{|C, nb >} $
	
	\vspace{2mm}
	
	$D_{n b}(t+1) \ \mathbf{|D, nb >} = \ \mathbf{\hat{W}} \ (A_{na}(t) s_{n} s_{1} \ - \ B_{na}(t) \ s_{n} c_{1} - C_{n b}(t)  c_{n} s_{1} + D_{n b}(t) \ c_{n} c_{1}) \ \mathbf{|D, nb >} $
	
	\vspace{2mm}
	
	\item	$\mathbf{2}$
	
	If now we let the operator $\mathbf{\hat{W}}$ to act
	
	\vspace{2mm}
	
	$A_{n a}(t+1) \mathbf{|A, na >} = \  (A_{na}(t) \ c_{n} c_{1} \ + \ B_{na}(t) \ c_{n} s_{1} + C_{n b}(t) \ s_{n} c_{1} + D_{n b}(t) \ s_{n} s_{1}) \ \mathbf{|A, na + a >} $
	
	\vspace{2mm}
	
	$B_{n a}(t+1) \mathbf{|B, na >}  = \  (- A_{na}(t) \ c_{n} s_{1} \ + \ B_{na}(t) \ c_{n} c_{1} - C_{n b}(t) \ s_{n} s_{1} + D_{n b}(t) \ s_{n} c_{1}) \ \mathbf{|B, na -a>} $
	
	\vspace{2mm}
	
	$C_{n b}(t+1) \mathbf{|C, nb >} = \ (- A_{na}(t) s_{n} c_{1} \ - \ B_{na}(t) \ s_{n} s_{1} + C_{n b}(t)  c_{n} c_{1} + D_{n b}(t) \ c_{n} s_{1}) \ \mathbf{|C, nb + b>} $
	
	\vspace{2mm}
	
	$D_{n b}(t+1) \mathbf{|D, nb >} = \ (A_{na}(t) s_{n} s_{1} \ - \ B_{na}(t) \ s_{n} c_{1} - C_{n b}(t)  c_{n} s_{1} + D_{n b}(t) \ c_{n} c_{1}) \ \mathbf{|D, nb - b>} $
	
	\vspace{4mm}
	
	\item $\mathbf{3}$
	Since we are given a magnetic field $B = B \ U_{z}$ \cite{QW B Fractal} , \cite{Feynman B} that goes through the axis of both rings, we must find some potential vector A such that $B = \nabla \times A$.
	\\
	
	And we know that the curl operator in polar coordinates is given by:
	\begin{equation}
	\begin{vmatrix}	
	U_{\rho} & U_{\Phi}  & U_{z}\\
	\partial_{\rho} &  1/ \rho \ \partial_{\Phi}  & \partial_{z} \\
	A_{\rho} & A_{\Phi}  &  A_{z}\\
	\end{vmatrix}	
	\end{equation}
	
	So a vector $\mathbf{A} = (0, R B, 0)$ fits with what we need.
	
	On the other hand and like is said by Feynman \cite{Feynman B} the amplitude $<b|a>_{B \neq 0} = <b|a>_{B=0} * \ exp(i \int \textbf{A} \textbf{ds})$ where A is the potential vector. And ds goes around a trajectory that encloses B
	\\
	
	Because we are dealing with a circunference $\mathbf{\vv{ds}} = R \ d\Phi \ \vv{U_{\Phi}}$. Furthermore is not the same thing going clockwise than going counterclockwise. Indeed we get this two different phases: $\mathbf{e^{ i \ \int_{\Phi}^{\Phi + \Delta \ \Phi} (R B) \vv{U_{\Phi}} * R d \Phi \vv{U_{\Phi}}}}$ for the first case, that leads us to $\mathbf{exp ( i \ R^{2} \ \textcolor{red}{\Phi} \ B) |_{\Phi}^{\Phi + \Delta \ \Phi}} $.
	\vspace{2mm}
	
	If we replace the integration limits this phase is obtained for the amplitude going clockwise
	\vspace{2mm}
	
	$exp \ ( i \ R^{2} \ (\textcolor{red}{(\Phi + \Delta \ \Phi) - \Phi})  \ B)$
	\vspace{2mm}
	
	And this one for the particle that goes counterclockwise
	\vspace{2mm}
	
	$exp \ ( i \ R^{2} \ (\textcolor{red}{(\Phi - \Delta \ \Phi) - \Phi})  \ B)$
	\vspace{2mm}
	
	So you must consider these two phases
	\begin{equation}
	exp \ ( i \ R^{2} \ \textcolor{red}{\Delta \ \Phi}  \ B)
	\end{equation}
	
	\begin{equation}
	exp \ ( - i \ R^{2} \ \textcolor{red}{\Delta \ \Phi}  \ B)
	\end{equation}
	
	And the following relations are obtained
	
	\vspace{2mm}
	
	$A_{n a}(t+1) \ \mathbf{|A, na >} = e^{( i \ R_{P}^{2} \ \textcolor{red}{ \Delta \ \Phi_{a}}  \ B)} \  (A_{na}(t) \ c_{n} c_{1} \ + \ B_{na}(t) \ c_{n} s_{1} + C_{n b}(t) \ s_{n} c_{1} + D_{n b}(t) \ s_{n} s_{1}) \ \mathbf{|A, na + a >} $
	
	\vspace{2mm}
	
	$B_{n a}(t+1) \ \mathbf{|B, na >}  = e^{(- i \ R_{P}^{2} \ \textcolor{red}{ \Delta \ \Phi_{a}}  \ B)} \  (- A_{na}(t) c_{n} s_{1} \ + \ B_{na}(t) \ c_{n} c_{1} - C_{n b}(t)  s_{n} s_{1} + D_{n b}(t) \ s_{n} c_{1}) \ \mathbf{|B, na -a>} $
	
	\vspace{2mm}
	
	$C_{n b}(t+1) \mathbf{|C, nb >} = e^{( i \ R_{G}^{2} \ \textcolor{red}{ \Delta \ \Phi_{b}}  \ B)} \ (- A_{na}(t) s_{n} c_{1} \ - \ B_{na}(t) \ s_{n} s_{1} + C_{n b}(t)  c_{n} c_{1} + D_{n b}(t) \ c_{n} s_{1}) \ \mathbf{|C, nb + b>} $
	
	\vspace{2mm}
	
	$D_{n b}(t+1) \mathbf{|D, nb >} = e^{(- i \ R_{G}^{2} \ \textcolor{red}{ \Delta \ \Phi_{b}}  \ B)} \ (A_{na}(t) s_{n} s_{1} \ - \ B_{na}(t) \ s_{n} c_{1} - C_{n b}(t)  c_{n} s_{1} + D_{n b}(t) \ c_{n} c_{1}) \ \mathbf{|D, nb - b>} $

	\vspace{4mm}
	
	\item	$\mathbf{4}$
	
	Right now, on the right hand of the equations we have just described, we do the next change of variable n $\longrightarrow$ n + 1 or n  $\longrightarrow$ n - 1 depending on the row we are. But here some caution must be taking into consideration, not only the ``n'' in the amplitudes changes but also does the ``n'' in $\Theta_{n}$
	
	\vspace{2mm}
	
	$A_{n a}(t+1) \ \mathbf{|A, na>} = e^{( i \ R_{P}^{2} \ \textcolor{red}{ \Delta \ \Phi_{a}}  \ B)} \  (A_{(n-1)*a}(t) \ c_{n-1} c_{1} \ + \ B_{(n-1)*a}(t) \ c_{n-1} s_{1} + C_{(n-1)*b}(t) \ s_{n-1} c_{1} + D_{(n-1)*b}(t) \ s_{n-1} s_{1}) \ \mathbf{|A, na>} $
	
	\vspace{2mm}
	
	$B_{n a}(t+1) \ \mathbf{|B, na>} = e^{( - i \ R_{P}^{2} \ \textcolor{red}{ \Delta \ \Phi_{a}}  \ B)} \  (- A_{(n+1)*a}(t) c_{n+1} s_{1} \ + \ B_{(n+1)*a}(t) \ c_{n+1} c_{1} - C_{(n+1)*b}(t)  s_{n+1} s_{1} + D_{(n+1)*b}(t) \ s_{n+1} c_{1}) \ \mathbf{|B, na>} $
	
	\vspace{2mm}
	
	$C_{n b}(t+1) \ \mathbf{|C, nb>} = e^{( i \ R_{G}^{2} \ \textcolor{red}{ \Delta \ \Phi_{b}}  \ B)} \ (- A_{(n-1)*a}(t) s_{n-1} c_{1} \ - \ B_{(n-1)*a}(t) \ s_{n-1} s_{1} + C_{(n-1)*b}(t)  c_{n-1} c_{1} + D_{(n-1)* b}(t) \ c_{n-1} s_{1}) \ \mathbf{|C, nb>} $
	
	\vspace{2mm}
	
	$D_{n b}(t+1) \ \mathbf{|D, nb>} = e^{( - i \ R_{G}^{2} \ \textcolor{red}{ \Delta \ \Phi_{b}}  \ B)} \ (A_{(n+1)*a}(t) s_{n+1} s_{1} \ - \ B_{(n+1)*a}(t) \ s_{n+1} c_{1} - C_{(n+1)*b}(t)  c_{n+1} s_{1} + D_{(n+1)*b}(t) \ c_{n+1} c_{1}) \ \mathbf{|D, nb>} $
	
	\vspace{4mm}
	
	These equations are the MAP of this theorethical approach for the interaction of two particles moving around two concentric quantum rings. They, of course, depend on two different variables $K_{a}$ for the outer ring, and $K_{b}$ for the inner one. this is how we obtain the set of spectrums we show in figures \textbf{3},\textbf{4}, and \textbf{5}. Firstly we consider the magnetic field applied B = 0,  and then we apply some B different from 0 and we observe a shifting in the spectrum. This last case represent an example of how the Aharonov-Bohm effect appears in our model
	
\end{itemize}

\begin{figure*}[htb]
\captionsetup{justification=centering}
\begin{subfigure}{0.3\textwidth}
	\includegraphics[width=\linewidth]{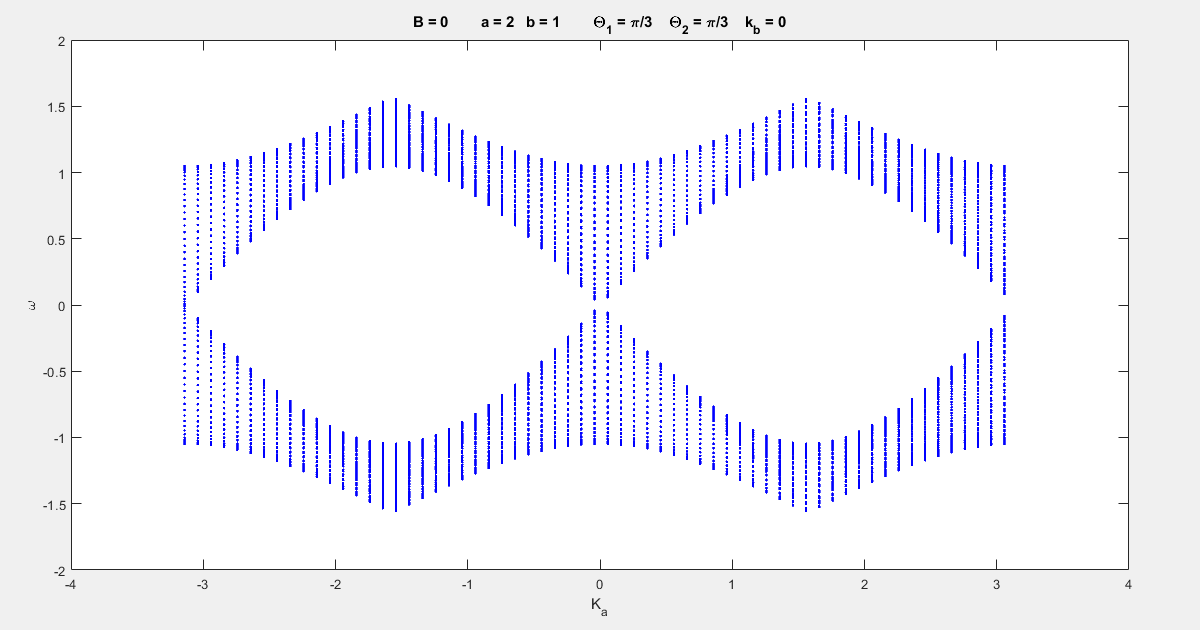}\quad
	\caption{a = 2, b = 1}
	\label{fig:1}
\end{subfigure}
\begin{subfigure}{0.3\textwidth}
	\includegraphics[width=\linewidth]{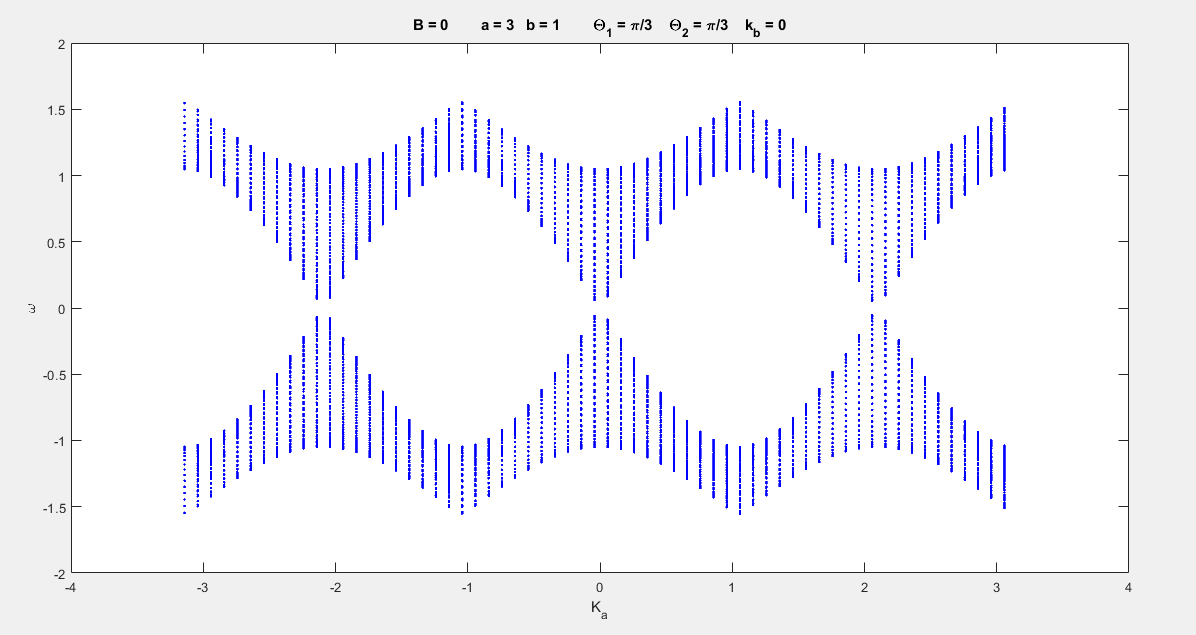}\quad
	\caption{a = 3, b = 1}
	\label{fig:2}
\end{subfigure}
\begin{subfigure}{0.3\textwidth}
	\includegraphics[width=\linewidth]{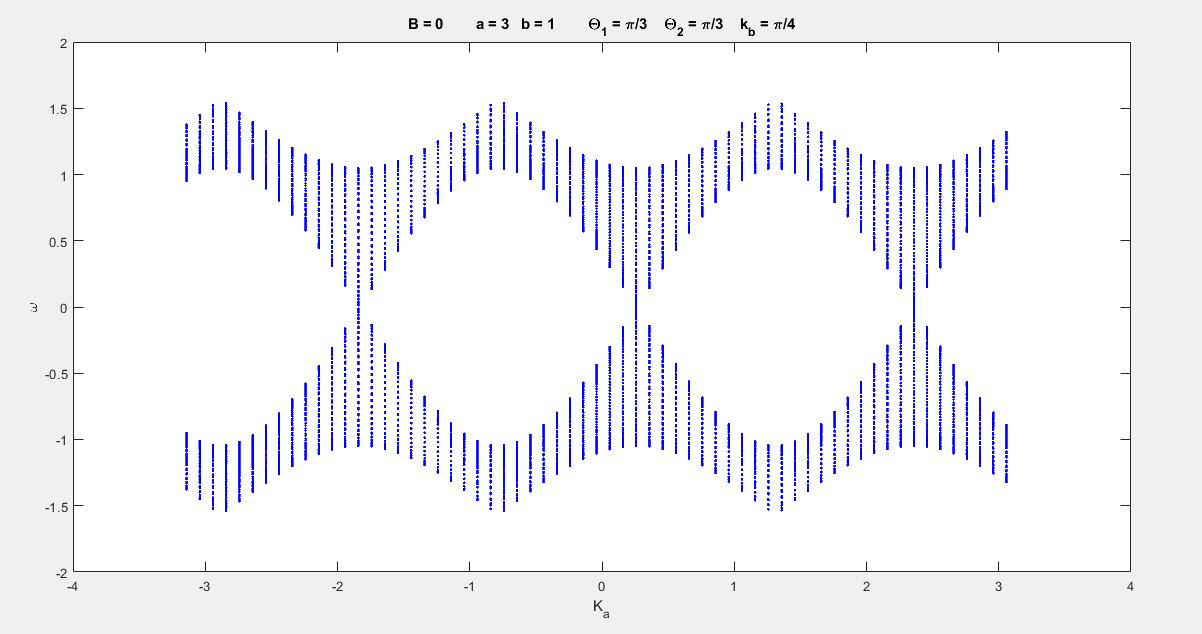}\quad
	\caption{a = 3 , b = 1 , $K_{b} \neq 0$}
	\label{fig:3}
\end{subfigure}

	\medskip
\begin{subfigure}{0.3\textwidth}
	\includegraphics[width=\linewidth]{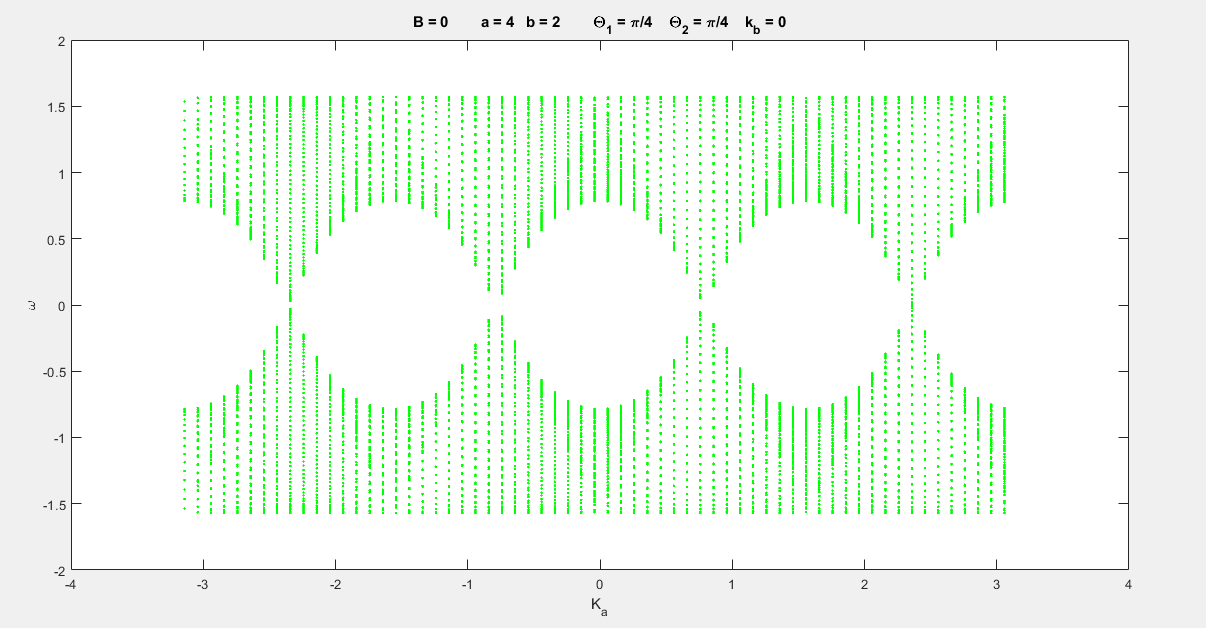}\quad
	\caption{a = 4 , b = 2}
	\label{fig:4}
\end{subfigure}
\begin{subfigure}{0.3\textwidth}
   \includegraphics[width=\linewidth]{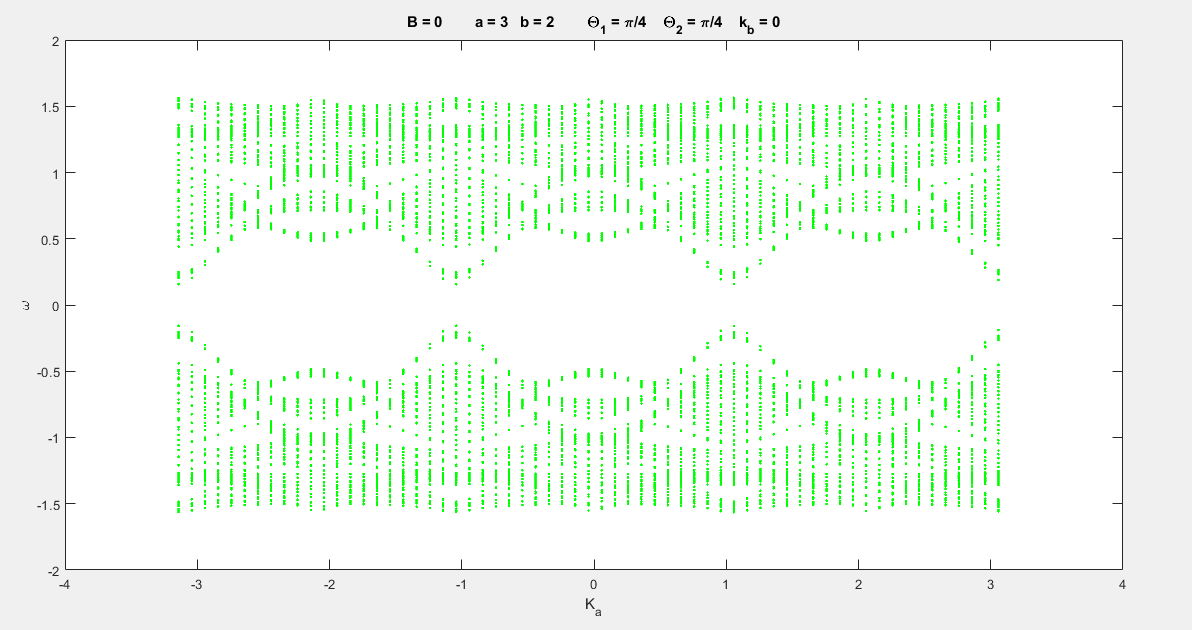}\quad
   \caption{a = 3 , b = 2}
   \label{fig:5}
\end{subfigure}
\begin{subfigure}{0.3\textwidth}
   \includegraphics[width=\linewidth]{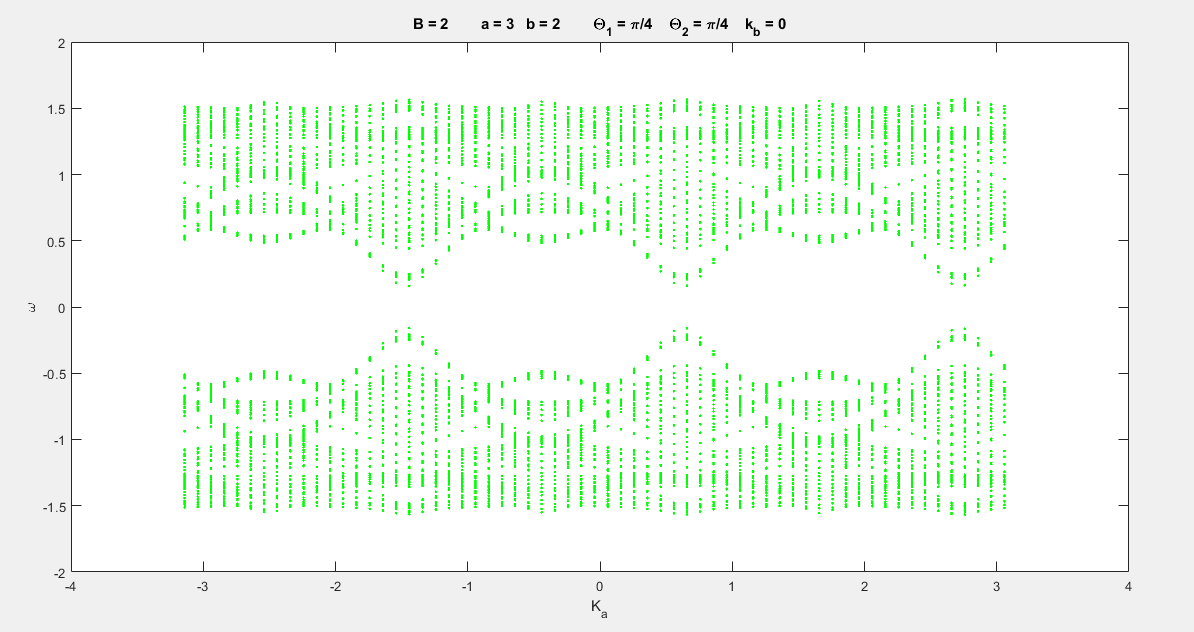}
   \caption{a = 3 , b = 2 , $B \neq 0$}
   \label{fig:6}
\end{subfigure}

    	\medskip
    \begin{subfigure}{0.3\textwidth}
    	\includegraphics[width=\linewidth]{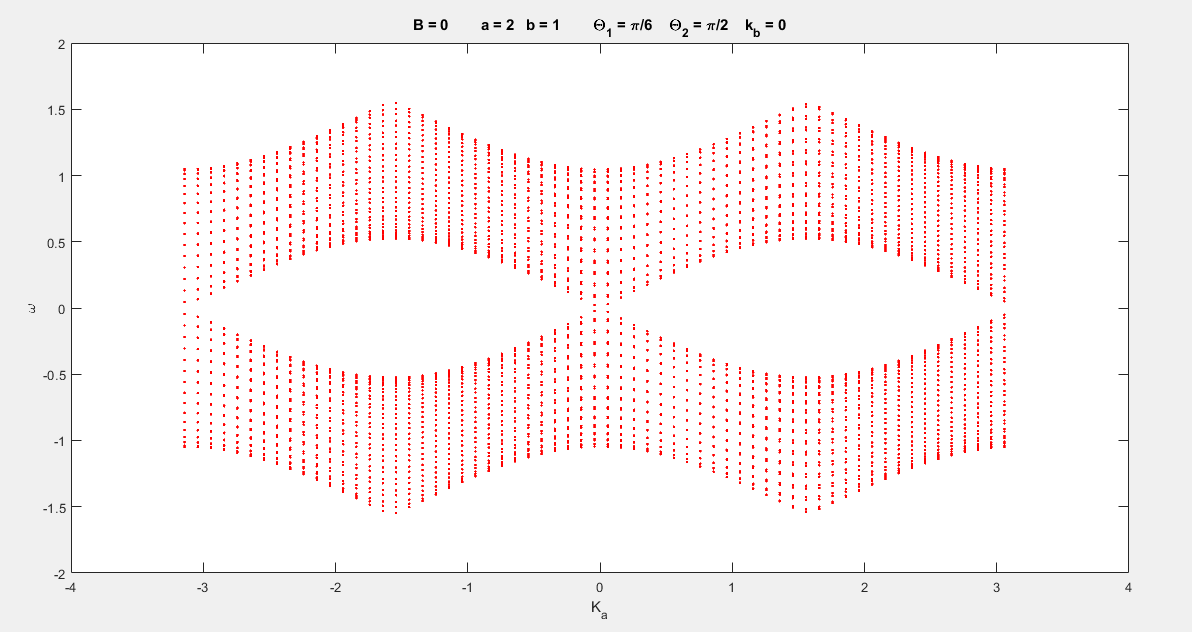}\quad
    	\caption{a = 2 , b = 1}
    	\label{fig:7}
    \end{subfigure}
    \begin{subfigure}{0.3\textwidth}
    	\includegraphics[width=\linewidth]{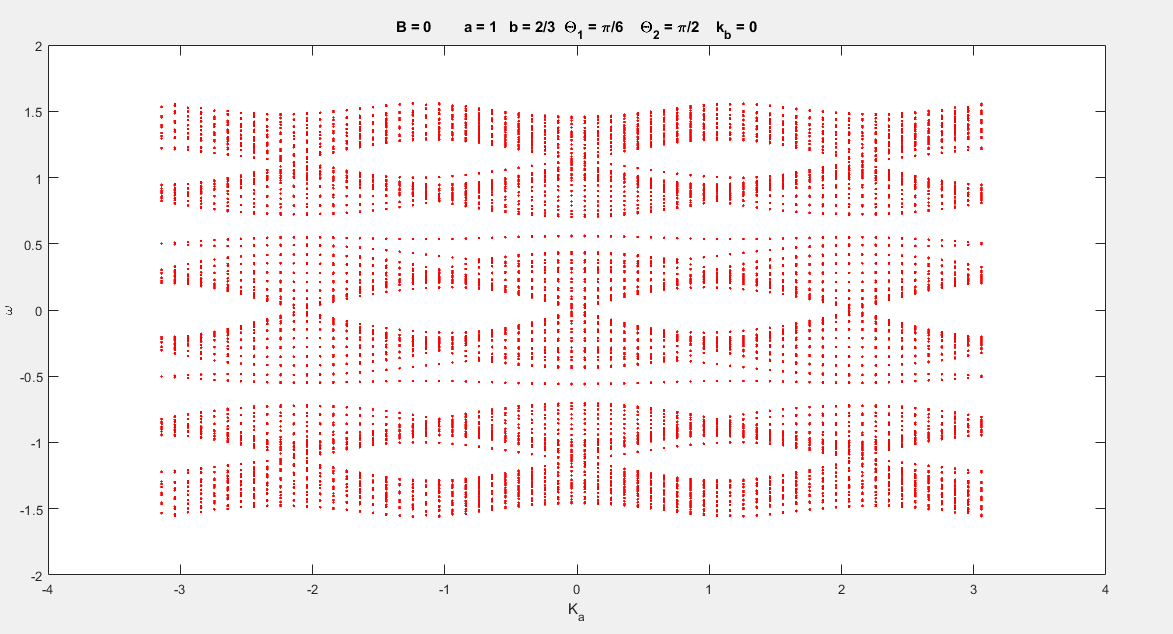}\quad
    	\caption{a = 1 , b = $2/3$}
    	\label{fig:8}
    \end{subfigure}
    \begin{subfigure}{0.3\textwidth}
    	\includegraphics[width=\linewidth]{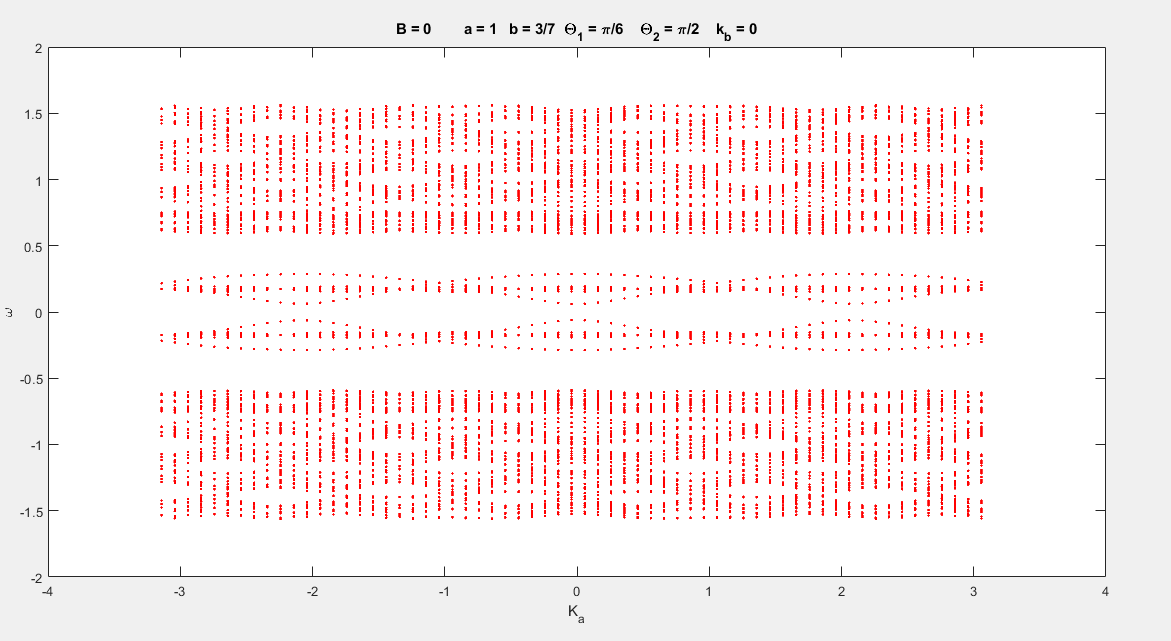}
    	\caption{a = 1 , b = $3/7$}
    	\label{fig:9}
    \end{subfigure}
\caption{Several spectra for some \textbf{QR} with a certain value of $K_{b}$ fixed. A quantity of N = 15 sites has been chosen. You should take notice of the different behaviour depending on whether or not a = n times b, with n an integer. So to speak, wheter or not an usual spectrum of a \textbf{QW} is obtained}

\end{figure*}

\begin{figure*}
	\begin{subfigure}{0.4\textwidth}
		\includegraphics[width=\linewidth]{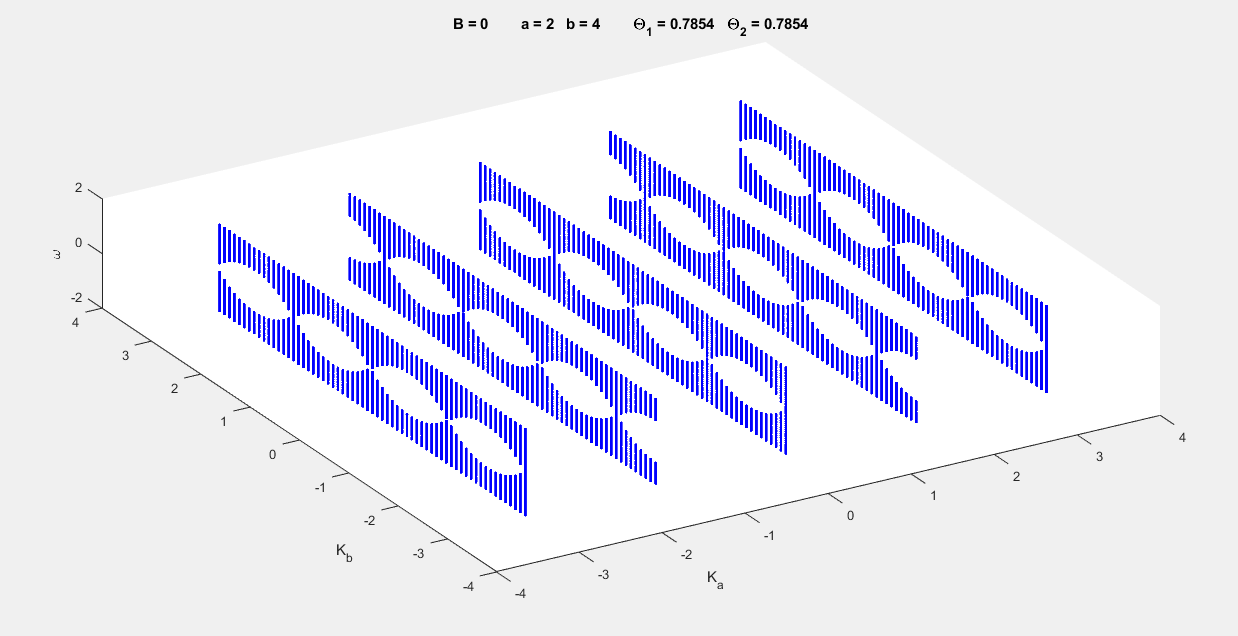}
		\caption{QR spectrum $K_{b}$}
	\end{subfigure}\hfil
    	\begin{subfigure}{0.4\textwidth}
    	\includegraphics[width=\linewidth]{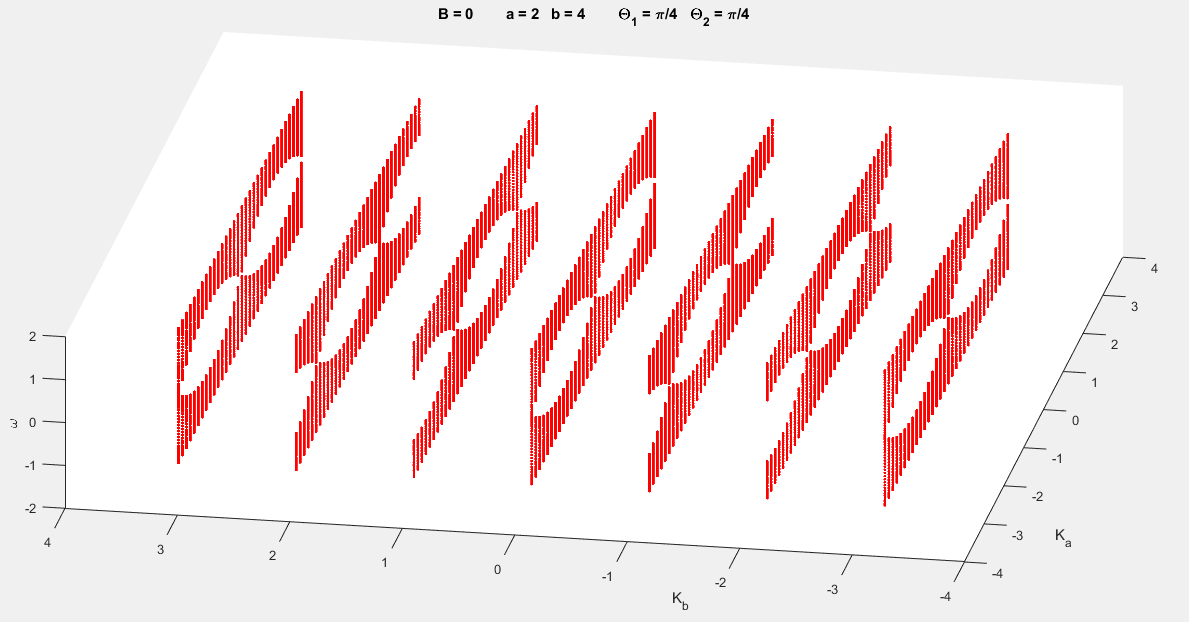}
    	\caption{QR spectrum $K_{a}$}
    \end{subfigure}
	\caption{These spectra depend on $K_{a}$ and $K_{b}$}
\end{figure*}

\begin{figure*}[h!]
	\captionsetup{justification=centering}
	\begin{subfigure}{0.45\textwidth}
		\includegraphics[width=\linewidth]{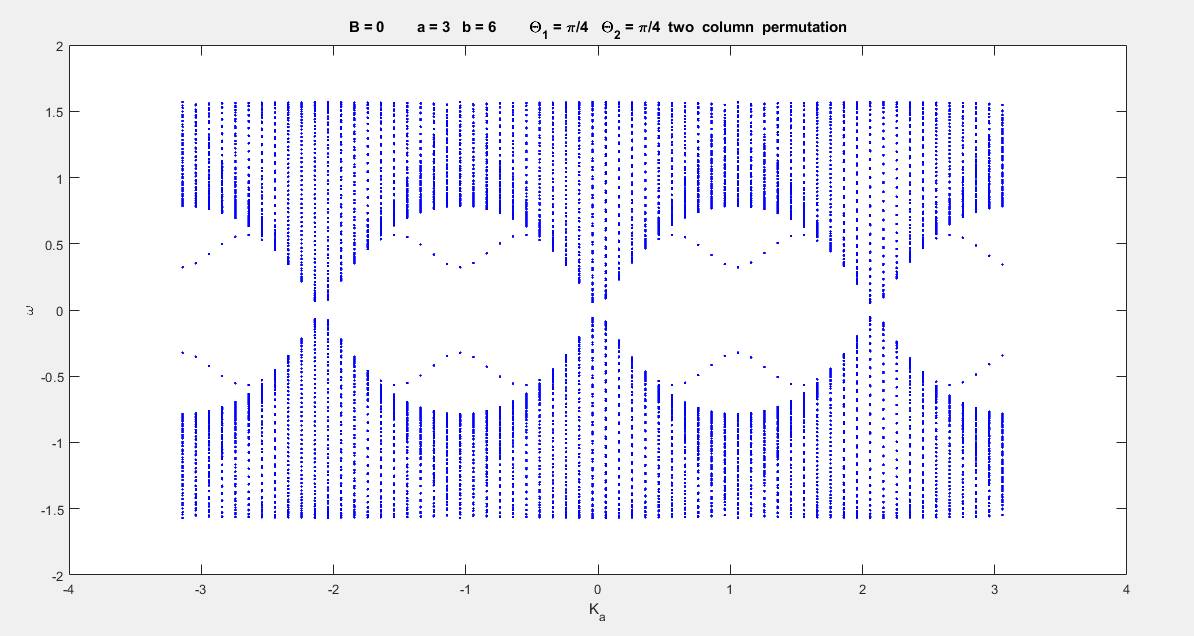}\quad
	\end{subfigure}\hfill	
	\begin{subfigure}{0.45\textwidth}
		\includegraphics[width=\linewidth]{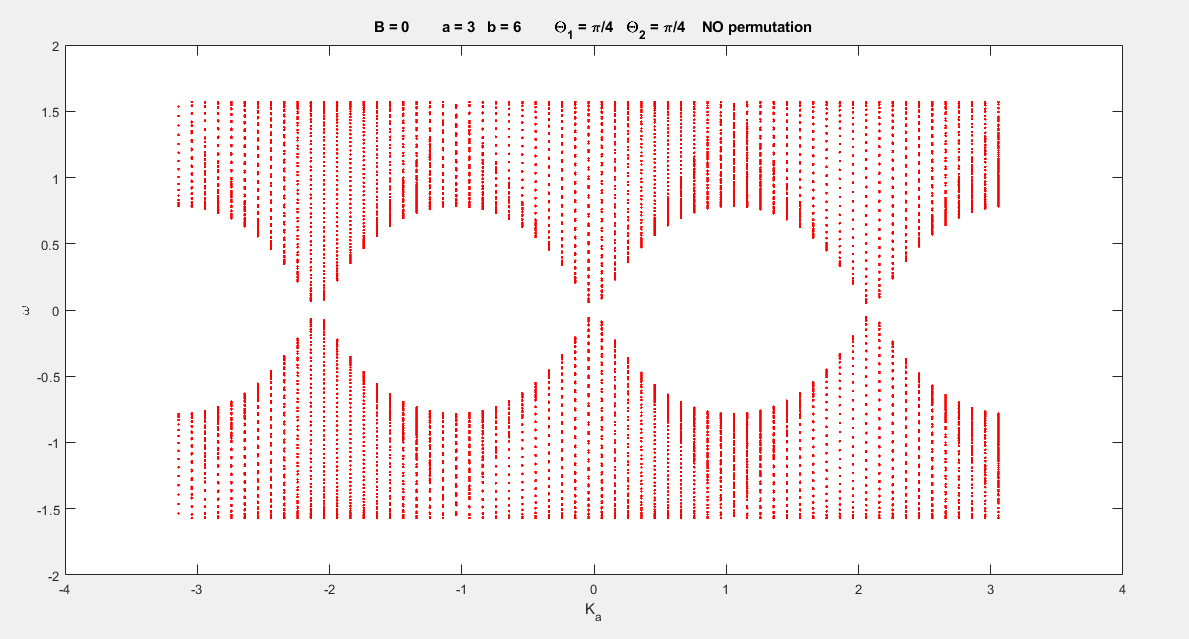}\quad
	\end{subfigure}
	\caption{Breaking of symmetry on the left and standard periodical lattice on the right}
\end{figure*}

\begin{figure*}[h!]
	\begin{subfigure}[b]{0.45\textwidth}
		\includegraphics[width=\linewidth]{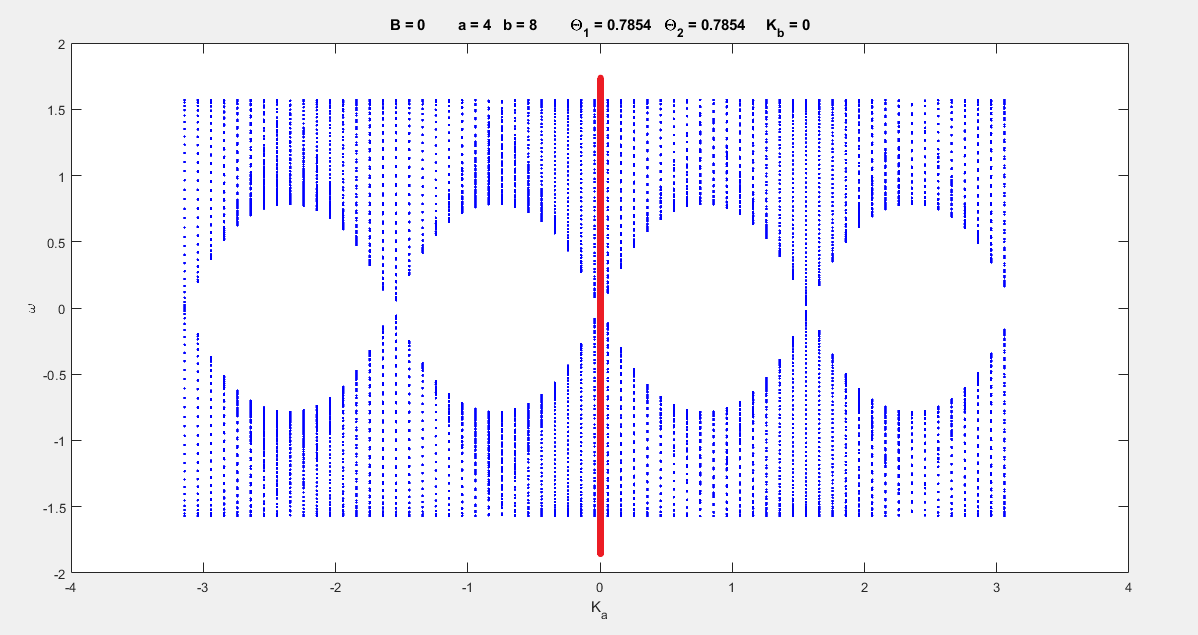}\quad
		\caption{B = 0}
	\end{subfigure}\hfill
    \begin{subfigure}[b]{0.45\textwidth}
    	\includegraphics[width=\linewidth]{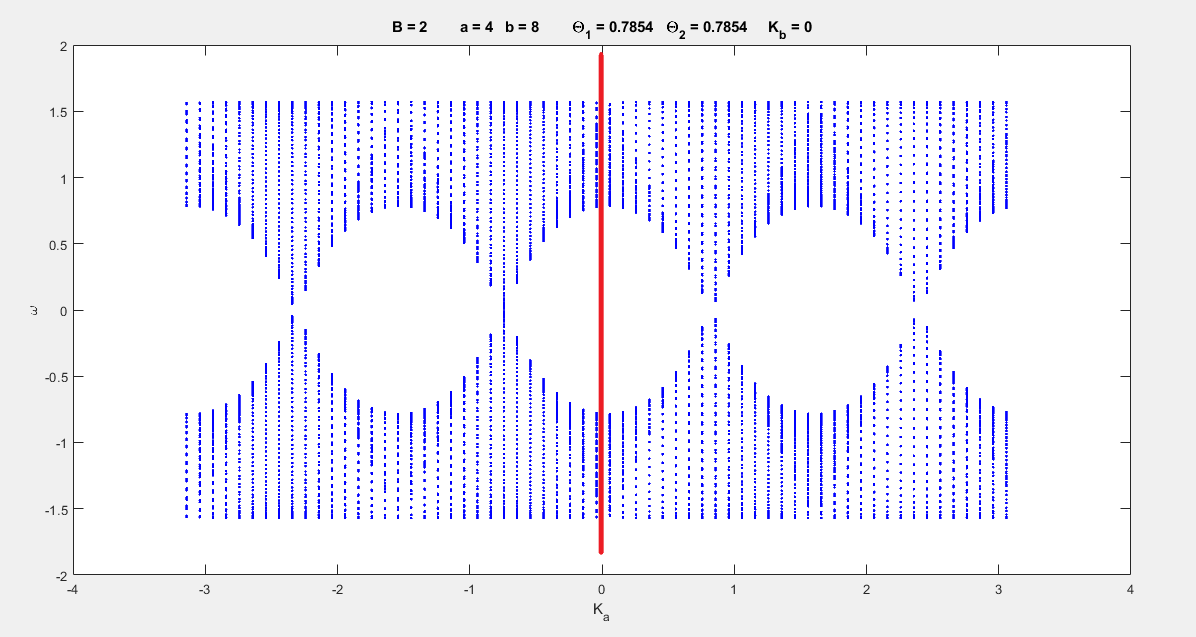}
    	\caption{B = 2}
    \end{subfigure}
\caption{ Magnetic Field applied. Before and after}
\end{figure*}	

In figure \textbf{3} nine examples of spectra are shown.The fist row correspond to $\Theta_{1} = \Theta_{2} = \pi/3$, the second one does to $\Theta_{1} = \Theta_{2} = \pi/4$. And in the third one we use $\Theta_{1} = \pi/6 \ \ \Theta_{2} = \pi/2$

Besides, in figure \textbf{4} some spectra are presented in such a way that they match to different values of $K_{a}$ and $K_{b}$

And finally, in figure \textbf{6} the effect of a magnetic field applied by the axis of the rings is witnessed. The shifting of the spectrum is clear, although the eigenvalues for each $K_{b}$ remain constant

\section{An unexpected breaking of symmetry}

It is well known that an unitary matrix $\mathbf{\hat{U}}$ is defined in such a way that $< \hat{U} \Psi \ | \ \hat{U} \Psi > = < \Psi| \Psi>$. To get this goal $U^{-1}$ must equal ${U}^{\dagger}$. And this is totally equivalent to say that all modulus squared of all the elements in all rows and columns must sum up to one. In our case the problem to solve is kind of similar, in the matrix by boxes shown in the left hand side of figure 6 if you focus on the column coloured in green or the row coloured in red you have that $M13 * M13^{\dagger} + M24 * M24^{\dagger} = Id$. Likewise if you look at the other matrix in the same image you have that $M1 * M1^{\dagger} + M24 * M24^{\dagger} + M3p * M3p^{\dagger}= Id$ if you look at the most left yellow column. Or you have that $M4G * M4G^{\dagger} + M3p * M3p^{\dagger} = Id$ if you focus on the row above

Here we face that situation, a situation in which we have a big matrix made up of smaller matrices. I mean by this that we are dealing with the spectrum matrix that is arranged in such a manner. This can be seen in figure \textbf{7}. In this case every row of submatrices multiplied all of them by their corresponding transposed conjugated must result in the identity matrix, and to get such a purpose we can interchange columns of matrices once we have ensured the bigger matrix is already unitary. In figure \textbf{7} left, all the matrices within the column in green can be interchanged with all the matrices in the column in yellow. If we do so we still have an unitary matrix, but the spectrum differs from the original one. These changes appear in the form of extra spectral curves that were not there before, but they are nothing special. Although they may seem molecules or some kind of bound state, when you look at the eigenstates you realize they are nothing but meaningless stuff. In left-hand side of figure \textbf{5} we can see an example of this effect.

The matrices that we have used to plot the pictures in figure \textbf{5 left} are constructed as follows :

\begin{widetext}
	
\begin{equation*}
\mathbf{M13} =
\begin{pmatrix}  
(& c_{1} * c_{n} & s_{1} * c_{n} & c_{1} * s_{n} & s_{1} * s_{n} \ ) \  \  exp(-i K_{a}) \\
& 0 & 0 & 0  &   0 \\
(& - s_{n} * c_{1}  & - s_{n} * s_{1}  & c_{n} * c_{1}   & c_{n} * s_{1} \ )  \ \ exp(-i K_{b})  \\
& 0  & 0  & 0  & 0    \\
\end{pmatrix}
\end{equation*}

\begin{equation*}
\mathbf{M24} =
\begin{pmatrix}
& 0 & 0 & 0  &   0 \\
(& - s_{1} * c_{n} & c_{1} * c_{n} & - s_{1} * s_{n} & c_{1} * s_{n} \ ) \ \  exp(i K_{a}) \\
& 0  & 0  & 0  & 0    \\
(&  s_{n} * s_{1}  & - s_{n} * c_{1}  & - c_{n} * s_{1}   & c_{n} * c_{1} \ )  \ \  exp(i K_{b})  \\
\end{pmatrix}
\end{equation*}

\end{widetext}

You must be aware that when you put one column in the place of another one, the exponential functions at the end of the rows of the matrix change their arguments. They are not any longer in the form $exp(i K_{a})$ but they appear in the form $exp(i (N+1) K_{a})$ where N stands for the amount of columns in between the two that were interchanged. This change is because from one point in the lattice you jump not to the next one, but to other point separated a certain distance away, that distance is just N


\begin{figure*}
		\captionsetup{justification=centering}
		\begin{subfigure}{0.45\textwidth}
	\includegraphics[width=\linewidth]{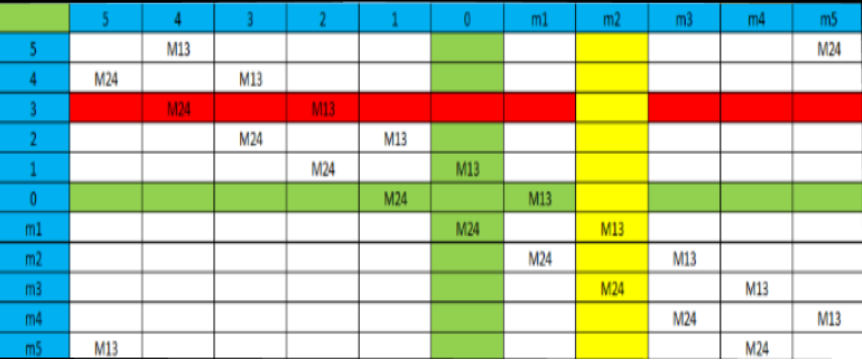}\quad
		\end{subfigure}\hfill	
		\begin{subfigure}{0.45\textwidth}
			\includegraphics[width=\linewidth]{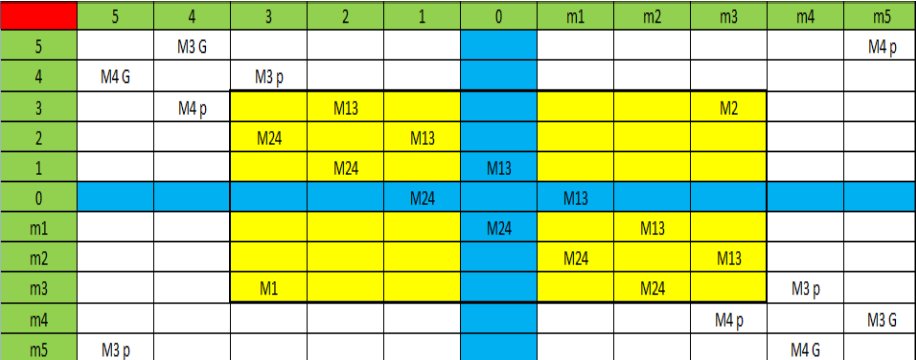}\quad
		\end{subfigure}
	\caption{Matrices of breaking of symmetry and of Moire pattern. We would like to emphasize that a matrix of boxes behaves in a similar way to a normal matrix, the rows and columns have to fulfil the same property in order for the global matrix to be unitary  }
\end{figure*}


\section{Quantum Propagation through a Moire pattern}

But this last section in which we emphasize a fact with seemingly little importance is what leads us towards a propagation through a Moire pattern. If we focused our attention on getting an unitary matrix such that replicates some kind of propagation over the geometry that is shown on top left of figure \textbf{9} we would be done.
\\

But it is not that easy to get such a matrix, not at all. In order to reproduce the foretold geometry we have to use a larger number of matrices for the construction of the spectrum than the ``usual'' matrices we used for the ``standard'' previous spectrum. We must engineer a spectrum such that any of its rows of smaller matrices behaves in the appropiate manner when we multiply them by the respective columns that appear in the transpose conjugated of the matrix of the spectrum itself. They must sum up to an identity matrix.
\\

This is maybe better understood looking at the matrices in figure \textbf{7}. These two matrices stand for the spectra of geommetric constructions similar to those shown in figure \textbf{1} and on top of figure \textbf{9}. Concerning the Moire pattern is the matrix on the right, the inner ``rectangle'' pretends to model the \textbf{QW} around the ring with a distance of four units between two consecutive sites. The dots in red in figure \textbf{9} left. And the other ring, represented by the circles in blue in the same picture with a separation of three units between nearest nodes, is somehow taken into account in the rest of the matrix. Because it has more sites. We would like to highlight that both the inner ``rectangle'' and the outer one are constructed to be cyclic because they both mimic two rings.
\\

The M13 and M24 boxes located in the inner ``rectangle'' were already extensively explained in the previous section. We introduce now some new matrices that are nothing but the same, except that they are somehow broken into pieces in order to achieve a global unitary matrix. We chase the same target of obtaining both rows and columns of smaller matrices in such a way that they sum up to a identity matrix after the multiplication by the transpose-conjugate matrices. We define these small matrices as follows:
\\

\begin{figure*}
	\captionsetup{justification=centering}
	\begin{subfigure}{0.45\textwidth}
		\includegraphics[width=\linewidth]{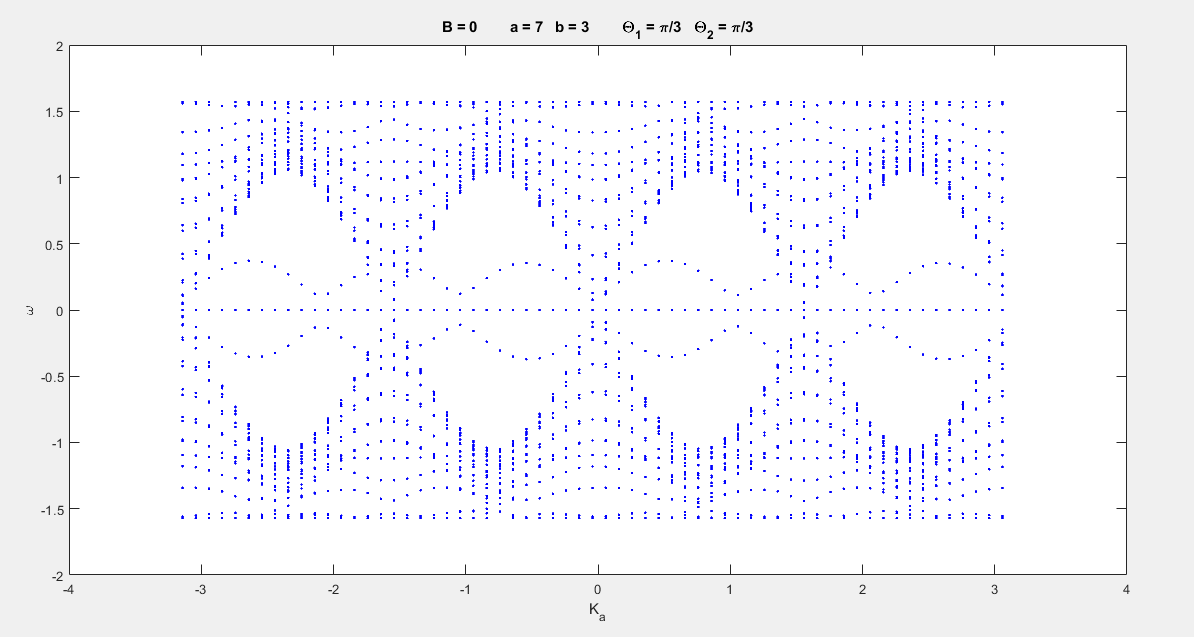}\quad
	\end{subfigure}\hfill	
	\begin{subfigure}{0.45\textwidth}
		\includegraphics[width=\linewidth]{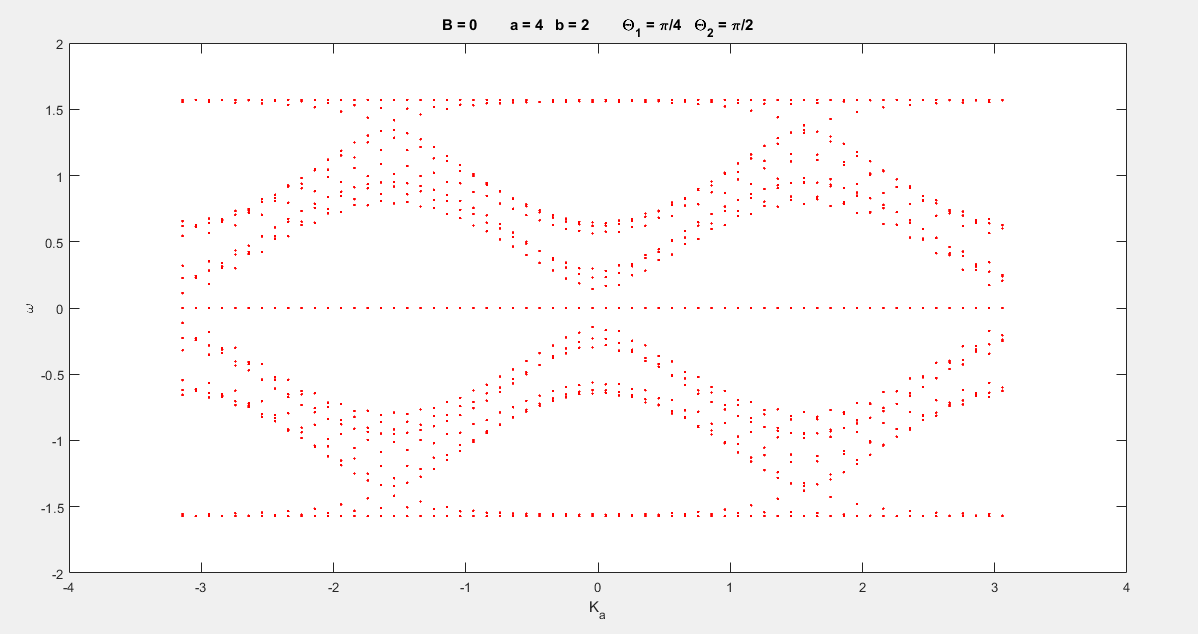}\quad
	\end{subfigure}
	\caption{2 spectra for a Moire pattern quantum ring}
\end{figure*}
\begin{widetext}
	
	\begin{equation*}
	\mathbf{M3p} =
	\begin{pmatrix} 
	   & 0 & 0 & 0 & 0 \\ & 0 & 0 & 0 & 0 \\
	   & 0 & 0 & c_{n} e^{-i \ K_{b}} & s_{n} e^{-i \ K_{b}} \\ & 0 & 0 & 0 & 0 
	\end{pmatrix}
%
	\ \ \ \ \mathbf{M3G} =
	\begin{pmatrix} 
	& 1 & 0 & 0 & 0 \\ & 0 & 1 & 0 & 0 \\
	& 0 & 0 & c_{n} e^{-i \ K_{b}} & s_{n} e^{-i \ K_{b}} \\ & 0 & 0 & 0 & 0 
	\end{pmatrix}
	\end{equation*}
	
	\begin{equation*}
	\mathbf{M4p} =
	\begin{pmatrix} 
	& 0 & 0 & 0 & 0 \\ & 0 & 0 & 0 & 0 \\ & 0 & 0 & 0 & 0 \\
	& 0 & 0 & - s_{n} e^{i \ K_{b}} & c_{n} e^{i \ K_{b}}
	\end{pmatrix}
%
	\ \ \ \ \mathbf{M4G} =
	\begin{pmatrix} 
	& 1 & 0 & 0 & 0 \\ & 0 & 1 & 0 & 0 \\ & 0 & 0 & 0 & 0 \\
	& 0 & 0 & - s_{n} e^{i \ K_{b}} & c_{n} e^{i \ K_{b}}
	\end{pmatrix}
	\end{equation*}
	
	\begin{equation*}
	\mathbf{M1} =
	\begin{pmatrix} 
	& c_{1} e^{-i \ K_{a}} & s_{1} e^{-i \ K_{a}} & 0 & 0 \\ & 0 & 0 & 0 & 0 \\ & 0 & 0 & 0 & 0 \\
	& 0 & 0 & 0 & 0 
	\end{pmatrix}
%
	\ \ \ \ \mathbf{M2} =
	\begin{pmatrix} 
	& 0 & 0 & 0 & 0 \\ & - s_{1} e^{i \ K_{a}}& c_{1} e^{i \ K_{a}}& 0 & 0  \\ & 0 & 0 & 0 & 0 \\
	& 0 & 0 & 0 & 0
	\end{pmatrix}
	\end{equation*}
	
\end{widetext}	
	

\begin{figure*}
	
		\begin{subfigure}{.45\textwidth}
			\includegraphics[width=\textwidth]{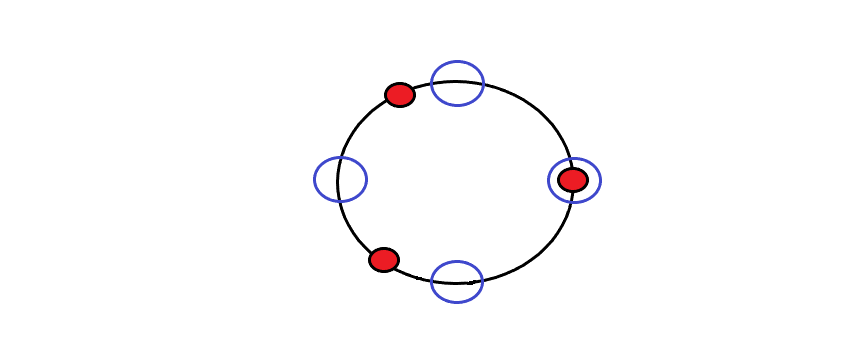}
			\caption{This Moire ring must be understood as two different rings. The first one, the one with blue dots is placed above the another one. They ``talk to each other'' only at one point}
		\end{subfigure}\hfill
		\begin{subfigure}{.45\textwidth}
			\includegraphics[width=\textwidth]{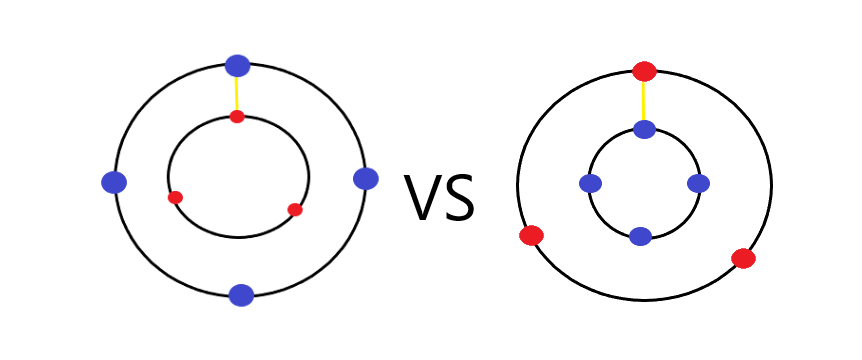}
			\caption{Two possibilities of almost equivalent Double Concentric Rings. Distances between nearest blue/red points differ from those in the previous frame}
		\end{subfigure}
		
		\vfill
		
		\begin{subfigure}{.35\textwidth}
			\includegraphics[width=\textwidth]{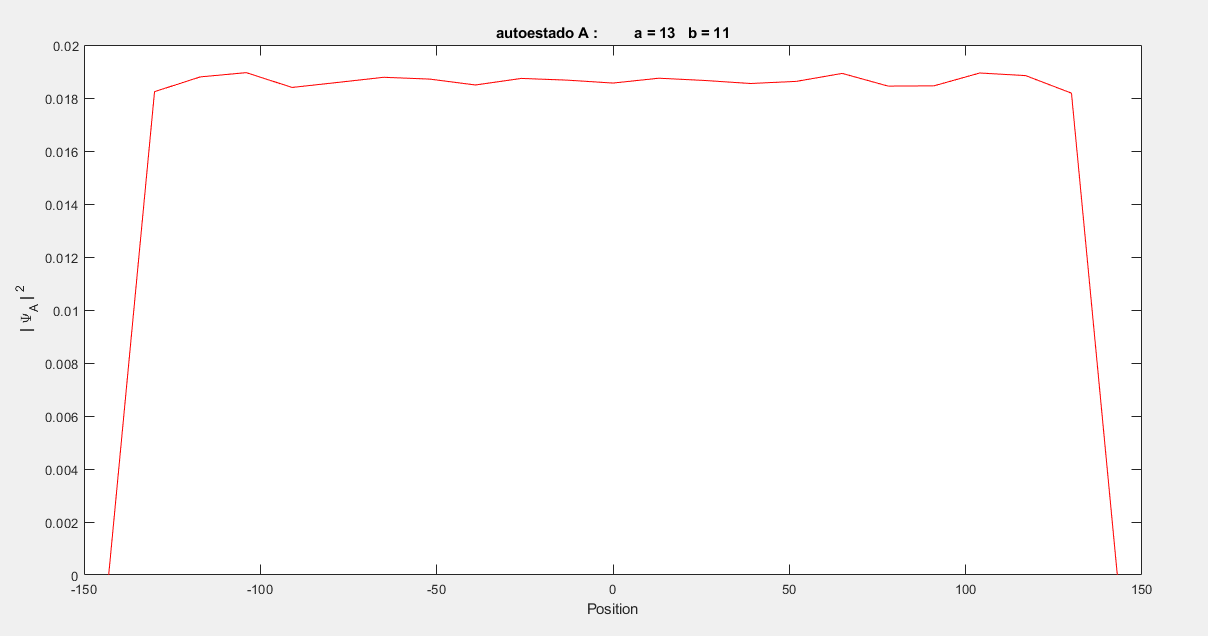}
			\caption{Prob. A, $N_{1} \neq N_{2}$}
		\end{subfigure}\hfill
		\begin{subfigure}{.35\textwidth}
			\includegraphics[width=\textwidth]{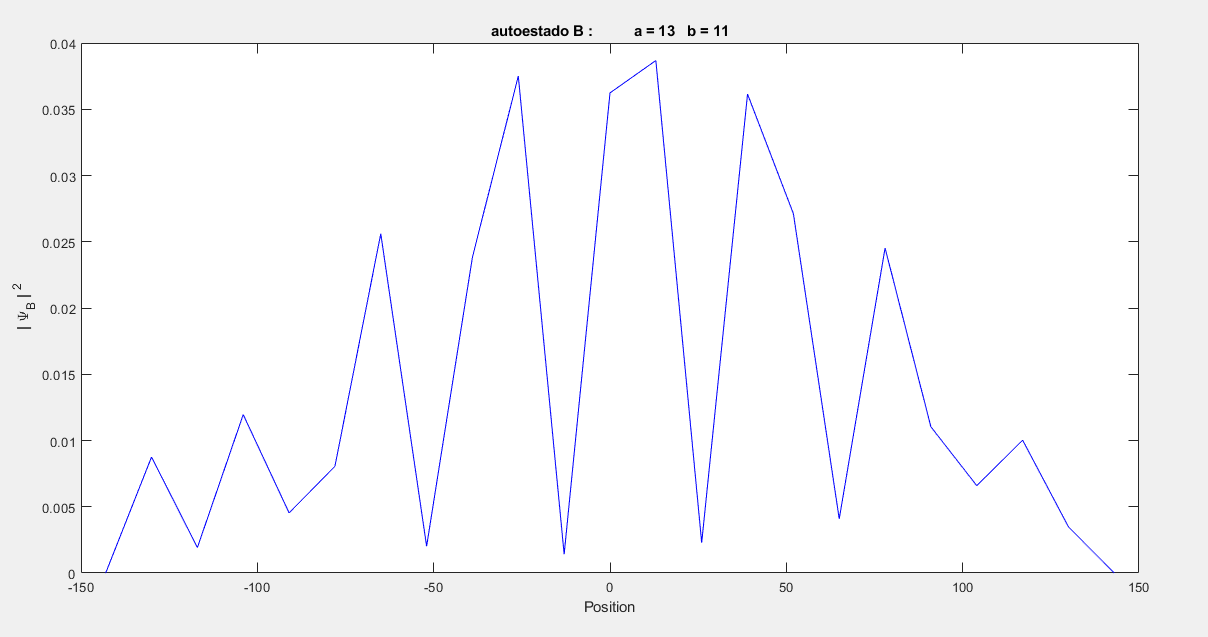}
			\caption{Prob. B, $N_{1} \neq N_{2}$}
		\end{subfigure}
		
		\vfill
		
		\begin{subfigure}[b]{.35\textwidth}
			\includegraphics[width=\textwidth]{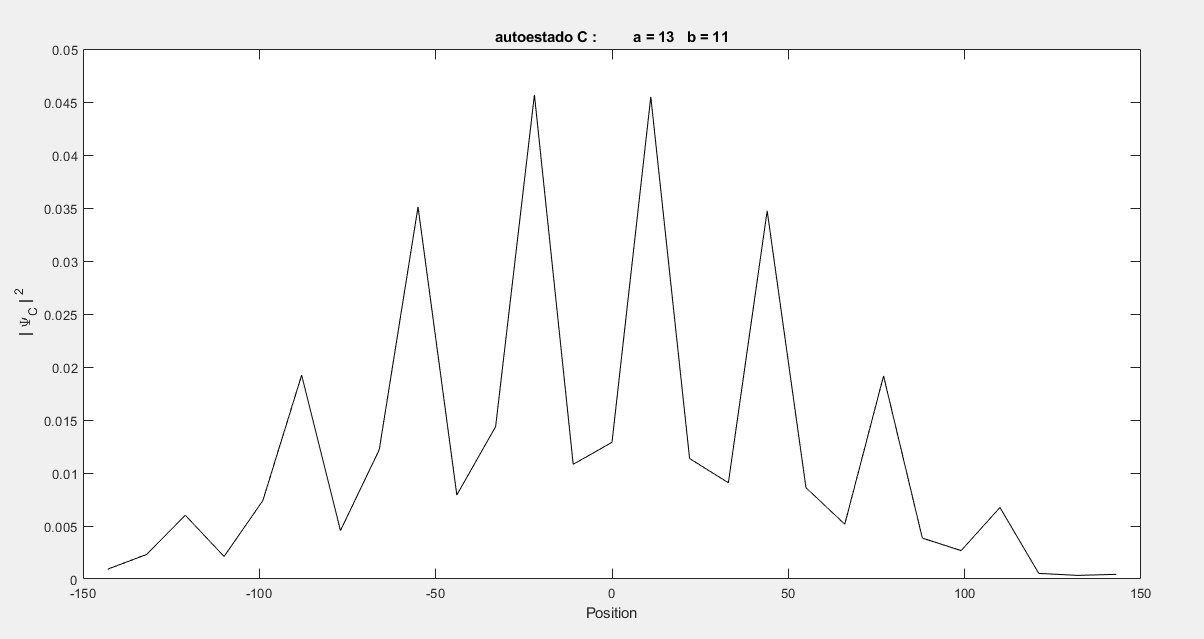}
			\caption{Prob. C, $N_{1} \neq N_{2}$}
		\end{subfigure}\hfill
		\begin{subfigure}[b]{.35\textwidth}
			\includegraphics[width=\textwidth]{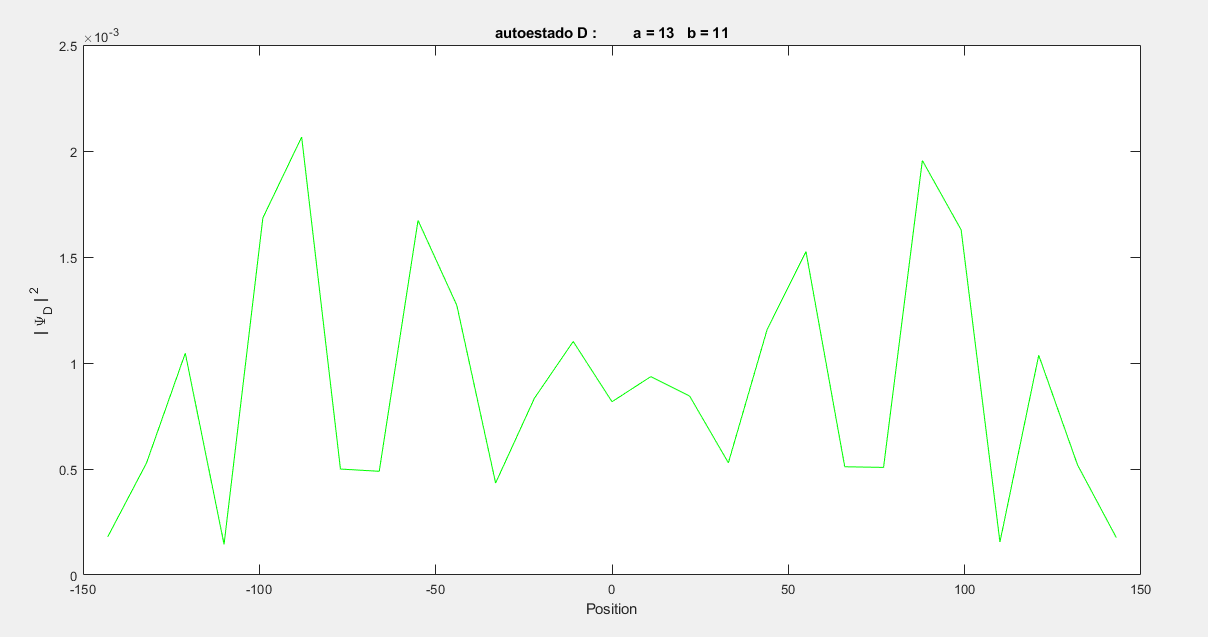}
			\caption{Prob. D, $N_{1} \neq N_{2}$}
		\end{subfigure}
    \caption{Moire \textbf{QW} schemes, and four distributions of probability for four random eigenstates}
\end{figure*}

  \begin{figure*}
  	\captionsetup{justification=centering}	
    \includegraphics[scale=0.4]{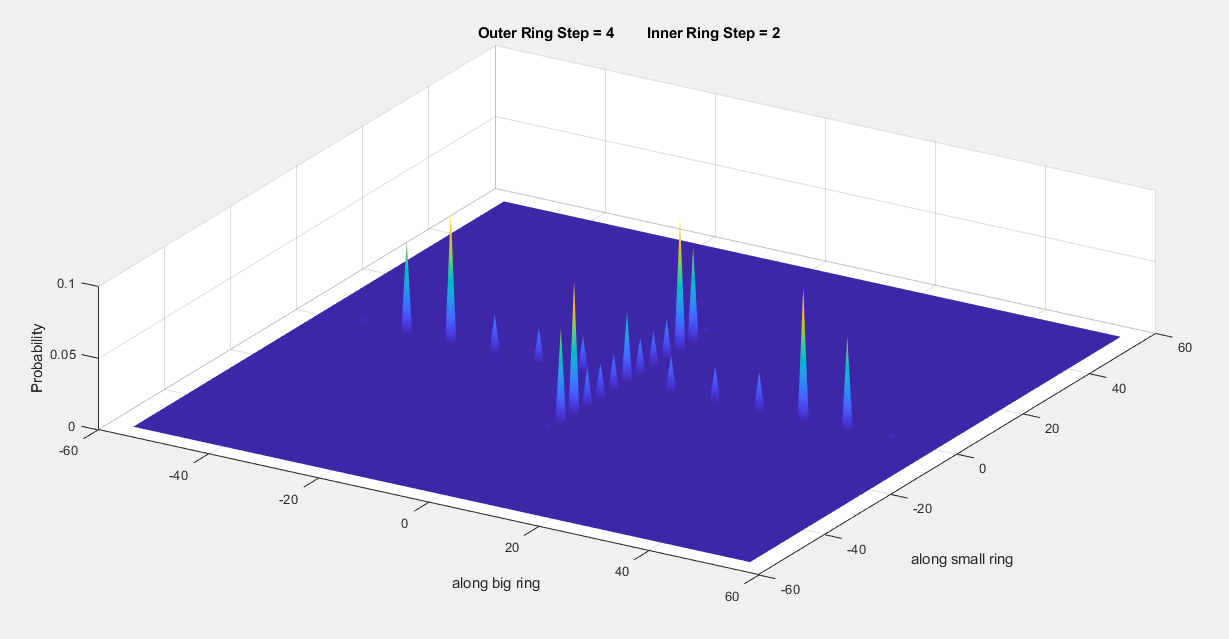}
    \caption{Propagation by a \textbf{QR}. Distribution of probability. Each \textbf{QW} represents a different particle}
  \end{figure*}  

\section{Conclusions}

$\bullet$ We have approached the problem of \textbf{QR} via \textbf{QW} and we have obtained a set of several spectra in which the frequency, hence the energy, depends on the wave number K, not on the angular momentum. We observe that, given $N_{1} = N_{2}$, situations in which the separation between consecutive nodes in one ring is an integer multiple of times the separation between nearest points in the another one lead to a spectrum with more peaks, the usual \textbf{QW} spectrum. And that translates into a higher group velocity of the moving particles. But we can not reproduce a wide range of phenomena already observed, even experimentally. A big instance of this limitation is the logic gate \cite{QR gate} achieved recently with the theory of angular momentum, \cite{Angular Momentum}
\\

$\bullet$ We have also applied a magnetic field by the axis of the two rings, and we have noticed a shifting in the spectrum. This is a kind of Aharonov-Bohm effect. It has nothing to do with what was done in \cite{QW B Fractal} 
\\

$\bullet$ We have engineered a possible way of tackling the Moire pattern problem, yet an operator should be encountered that makes possible the aforementioned \textbf{QW} in a more elegant manner.
\\

$\bullet$ Let me be clear on this. The angle $\Theta_{n}$ was defined in section \textbf{II} when we stablished the operator that models the distribution of the two electrons through the double concentric quantum rings in order to somehow mimic the tunnel effect. But there are much better forms of modeling such an angle. For example you could calculate the probability of tunneling between those rings taking into account the separation between them and define $\Theta_{n}$ according to that. Or you could also consider that the two rings are not perfectly aligned site by site, this implies $N_{1} \neq N_{2}$,  and  new Coin and Walk operators should be defined in such a case. Nevertheless We made the aforementioned definition of $\Theta_{n}$ in order to generate, in section \textbf{IV}, a Moire quantum walk without a lot of extra reshaping of the operators developed so far. \textit{Regarding this, one Moire Quantum Walk by a circle and a Quantum Walk by two double concentric rings with different number of points in each ring, both of them can have a very similar spectrum for two electrons propagating. Although some parameters are different. See figures \textbf{7b} and top of \textbf{9} for a better understanding}
\\

$\bullet$ We have checked out that the possibility for the two propagating particles to bind in a molecule is distint from zero in some cases in the double concentric quantum rings with the same number of sites in each one of the rings. The bigger the \textbf{lcm}(a,b) is, the more likelihood of obtaining a molecule there is. But such a probability turned out to be zero in the Moire model. So the framework developed here is very far from explaining the superconductivity found in \cite{MIT}. With only free propagating particles without interaction of any kind between them it seems impossible to succeed in this regard

\end{document}